# Demonstration of a-Si metalenses on a 12-inch glass wafer by CMOS-compatible technology

Ting Hu\* †, Qize Zhong†, Nanxi Li, Yuan Dong, Yuan Hsing Fu, Zhengji Xu, Dongdong Li, Vladimir Bliznetsov, Keng Heng Lai, Shiyang Zhu, Qunying Lin, Yuandong Gu, Navab Singh, and Dim-Lee Kwong

Institute of Microelectronics, Agency for Science, Technology and Research (A*STAR), 2 Fusionopolis Way, #08-02, Innovis, Singapore 138634, Singapore
†These authors contributed equally to this work
*hut@ime.a-star.edu.sg

## Abstract

Metalenses built up by artificial sub-wavelength nanostructures have shown the capability of realizing light focusing with miniature lens size. To date, most of the reported metalenses were patterned using electron beam lithography (EBL), which requires long processing time and is not suitable for mass production. Here, we demonstrate an amorphous silicon (a-Si) metalens on a 12-inch glass wafer via the 193 nm ArF deep UV immersion lithography, with critical dimension (CD) as small as 100 nm. The layer transfer technology is developed to solve the glass wafer handling issue in complementary metal-oxide-semiconductor (CMOS) fabrication line. The measured numerical aperture (NA) is 0.494 with a beam spot size of 1.26 µm, which agrees well with the simulation results. The focusing efficiency of 29.2% is observed at the designed wavelength of 940 nm. In addition, the metalens is applied in an imaging system, which further verifies its focusing functionality.

## 1. Introduction

Metasurfaces composed of quasi-periodic sub-wavelength nanostructures have attracted a lot of research interests recently since they can be artificially designed to tailor the wavefront of light in the sub-wavelength scale. By engineering the phase and the amplitude of the electromagnetic wave, miniature-sized flat optics components with various functionalities have been demonstrated [1,2]. Among these components, metalens has drawn significant attention due to its wide applications. Metalenses based on plasmonic resonance in nano metallic structures were firstly demonstrated to work in the visible and near-infrared wavelength regime [3-5]. Though the metallic antenna can be made as thin as 30 nm, the focusing efficiency is quite low due to the high optical loss. In order to improve the lens performance, researchers gradually move the focus to dielectric metasurfaces [6]. The selection of dielectric materials for metalenses are mainly based on two aspects: 1) the materials have low optical loss at the designed wavelengths, 2) the materials with higher refractive index are preferred since they can implement antennas with lower height and consequently alleviate the challenges in fabrication (lower aspect ratio for the etching process). Conventionally, the amorphous silicon (a-Si) is used for metalens working at the near-infrared and mid-infrared wavelength regime due to its low optical loss [7-12], while for the visible wavelength regime, titanium dioxide ($TiO_2$), gallium nitride (GaN) and silicon nitride (SiN) are appropriate materials [13-19]. There are a few works choosing silicon (a-Si or poly silicon) as the material for metalenses at the visible wavelength [20,21], but these lenses suffer from low focusing efficiency even with a very thin antenna layer. Furthermore, except the work reported in [12], to the best of our knowledge, all the rest metalenses demonstrated so far are patterned using the electron beam lithography (EBL), which requires long patterning time and hence is not suitable for mass production. In the meanwhile, for the practical applications, like in cameras [10], augmented reality (AR)/virtual reality (VR) devices [22], *etc.*, manufacturing of metalenses in large scale is in high demand. Although in [12] the UV lithography was used to pattern the matelenses on a 4-inch quartz wafer, the limited resolution hindered it from being used for shorter wavelength below 1 µm.

In this paper, we demonstrate an a-Si metalens working at 940 nm fabricated on a 12-inch glass wafer. The 193 nm ArF deep UV (DUV) immersion lithography is used rather than EBL to pattern our design with CD as small as 100 nm. Different from our earlier work [23] based on a 12-inch silicon wafer, the glass wafer is used in this work, which is hardly to be detected or handled by litho and etch tools. In order to solve this issue, a layer transfer process is developed for metalens fabrication. The performance of the fabricated metalens is characterized by measuring its NA, the focal-spot size and the focusing efficiency. The experimental results agree well with the 3D finite-difference time-domain (FDTD) simulation. Our work shows the feasibility for the mass production of the dielectric metalenses using the CMOS-compatible technology.

## 2. Device design and simulation

The schematic structure of the metalens design is shown in Fig. 1(a). It is constructed by sub-wavelength sized a-Si cylindrical pillars fabricated on a glass substrate and embedded within an interlayer. The interlayer, with a thickness of 60 µm, is induced during the layer transfer process to solve the glass wafer handling issue mentioned earlier. On top of the Si pillars, there is a 1-µm thick silicon dioxide (SiO2) cap layer. Fig. 1(b) and 1(c) shows side and top view of the meta-element respectively. The a-Si pillar height is 600 nm, with a fixed edge-to-edge gap of 200 nm to its neighboring pillars. To focus the collimated incident light, the relative phase profile of the meta-elements needs to follow the spatial distribution below [15]:

$$\varphi(x,y) = 2\pi - \frac{2\pi}{\lambda}\left(\sqrt{x^2 + y^2 + f^2} - f\right), \tag{1}$$

where λ is the designed wavelength and f is the focal length. The position of a-Si pillars is described by x and y. The phase shift of each building block can be adjusted by changing the diameter of the pillar. Through the 3D-FDTD simulation, the phase shift and the transmission of the a-Si pillar with respect to its diameter are shown in Fig. 1(d). Starting from the 100-nm diameter, the relative phase increases with larger diameter and reaches 2π at 294 nm. In the meanwhile, the transmission is higher than 0.94 when the diameter is smaller than 290 nm. From 290 nm to 294 nm, there is a drop of the transmission caused by the resonance in the pillars. While it should not have a significant effect on the metalens performance since 4-nm range is small compared with the whole diameter range from 100 nm to 294 nm. In the simulation, the optical wavelength is set to be 940 nm. The height of the a-Si pillars is chosen to be 600 nm, and the refractive index of the interlayer and a-Si at 940 nm are set as 1.451 (the same with the cap SiO$_2$ layer) and 3.806, respectively.

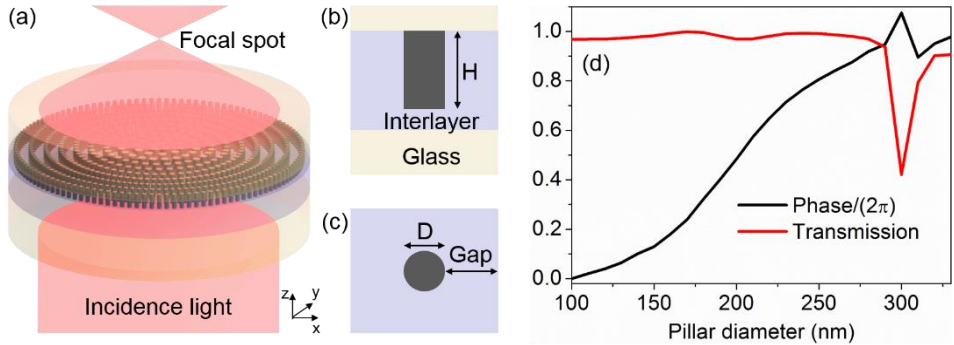

Fig. 1 (a) Schematic of the metalens. (b) The side view and (c) the top view of the metalens unit cell. (d) The simulated phase shift and transmission profile with respect to the pillar diameter.

Based on the phase shift obtained from the simulation [Fig. 1(d)] together with the spatial distribution of the phase profile described in Eq. (1), the a-Si pillars are placed accordingly to construct a metalens in the 3D-FDTD simulation. The original designed diameter of the metalens is 2 mm, but the 3D-FDTD simulation of such a large size requires huge computing resource. Therefore, the diameter of the metalens is scaled-down to be 20 µm with the scaling factor of 100 in simulation. Such simulation is setup as an analogy to investigate the performance of the original designed metalens. For the NA of 0.5, the simulated metalens is expected to have a focal length of 17.32 µm. The electric field intensity (E-intensity) in the light propagation plane (xz plane) and the E-intensity at the focal plane (xy plane) are shown Fig. 2(a) and 2(b), respectively. In Fig. 2(a), the spatial convergence of the incident plane wave can be observed after passing through the metalens located at z=0 µm. The incident wave is focused at z=17.32 µm as expected for the NA of 0.5. Fig. 2(b) shows the E-intensity distribution at the cross section of the focal plane, from which the focusing spot size and the efficiency can be obtained. Here the beam spot size is defined as the full width at half maximum (FWHM) of the E-intensity, and the focusing efficiency is defined by taking the ratio between the optical power within the circular area (diameter=4×FWHM) at the focal spot and the incident optical power on the metalens. In addition, 4 metalenses with different focal lengths of 4.843, 10, 17.32 and 31.798 µm, corresponding to the NA values of 0.9, 0.7, 0.5 and 0.3, respectively, are simulated. The focal spot size and the focusing efficiency are recorded, as shown in Fig. 2(c). The focal spot size is found to be increasing as the focal length increases. Its value is close to $\frac{\lambda}{2 \cdot NA}$, which indicates the focusing approaches the diffraction limit. The focusing efficiency has significant increasing from 39.6% to 76.2% as the focal length increases from 4.843 to 31.798 µm, which conveys that the bending of the incident wave with larger angle (higher NA) induces higher loss. The trade-off between the focusing efficiency and the NA should be considered in metalens design based on the application requirements.

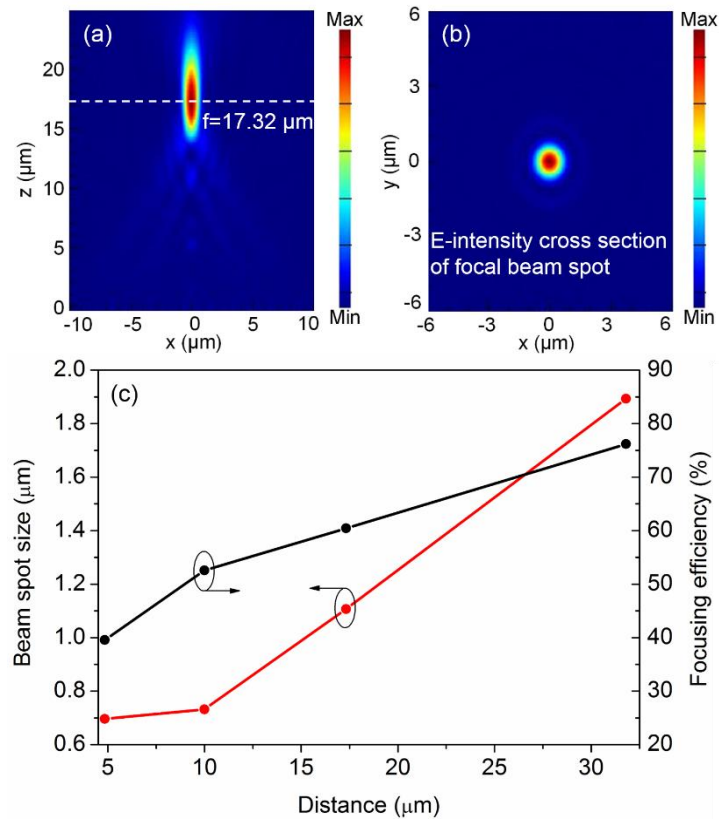

Fig. 2 The simulated E-intensity profile in (a) the propagation plane and (b) the focal plane of a 20-µm diameter metalens with a focal length of 17.32 µm (NA=0.5). (c) The simulated focal spot size and the focusing efficiency with respect to different focal lengths.

## 3. Fabrication and characterization

In the layout, a 2 mm-diameter metalens with the focal length of 1.732 mm (NA=0.5) was designed. The fabrication started with a standard 12-inch silicon wafer. A 1 µm-thick $SiO_2$ and a 600 nm-thick a-Si were deposited successively using the plasma-enhanced chemical vapor deposition (PECVD) method. Then the 193 nm ArF DUV immersion lithography and inductively coupled plasma (ICP) etch were used to make the a-Si pillars of the metalens. This was followed by an in-house developed layer transfer process which finally realizes metalenses on 12 inch glass wafer. Fig. 3(a) shows the picture of the fabricated wafer held by hands. Most of the area is without pattern and hence transparent, through which the A*STAR logo (blue color) placed at the back can be visualized. Fig. 3(b) and 3(c) show the top-view scanning electron microscope (SEM) images of the central and near-central zones of the metalens, respectively. Fig. 3(d) shows a zoomed-in angled-view SEM image of the pillars close to the edge of the metalens. The diameters of these pillars change from large to small values to cover the phase shift from 0 to 2π. All the pillars in this image, including the ones with 100 nm diameter (the 2nd column counting from right), have straight and smooth side wall. It verifies the quality of lithography and etching process.

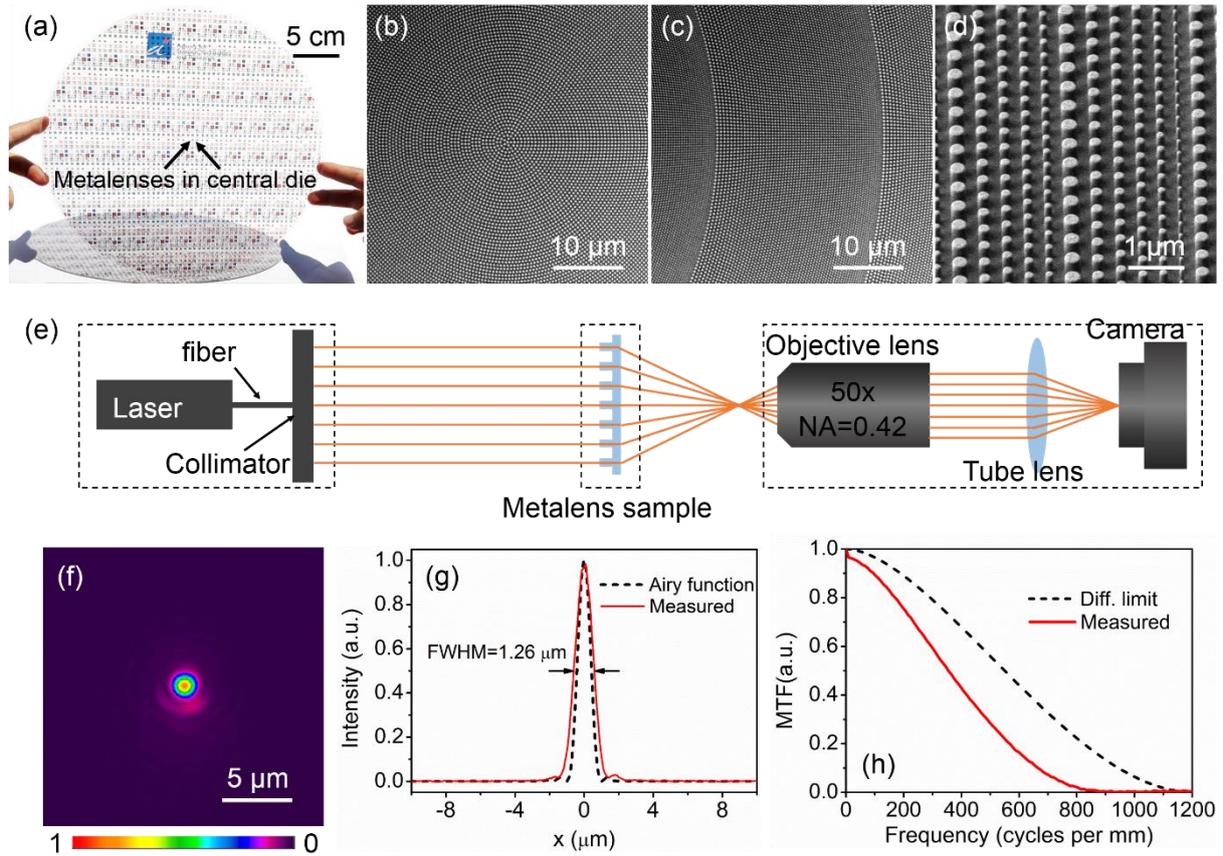

Fig. 3 (a) Color photo of the fabricated 12-inch glass wafer. SEM images at (b) central, (c) near-central and (d) outer zone of the metalens. (e) Schematic of the metalens characterization setup (not to scale). (f) Measured focal spots of the metalens at the design wavelength of 940 nm. (g) Horizontal cut of the focal spot shown in f. An ideal Airy function with the aperture of 2 mm and the focal length of 1.76mm is overlaid. (h) modulation transfer function (MTF) calculated using the spot profile shown in g. The dashed black line represents the diffraction limited MTF with the aperture diameter of 2 mm and the focal length of 1.76mm.

The schematic of the metalens characterization setup is shown in Fig. 3(e). The 940 nm laser diode has a fiber pigtail connected to a fiber collimator. The collimated light, with a Gaussian intensity profile, incidents onto the metaslens and then converges to a focal spot, which is later magnified by the objective lens (NA=0.42, 50× magnification) and the tube lens. A camera is used at the end to capture the magnified focal spot. The intensity distribution of the focal spot is shown in Fig. 3(f). The focal length is measured to be 1.76 mm, corresponding to an NA of 0.494, which matches with the designed value of 0.5. Based on the definition mentioned in the device design part, the focusing efficiency of the metalens is measured to be 29.2%, with the wafer transmission loss calibrated out. It is worth to mention that the metalens has a higher NA than the objective lens used in the system, which causes additional loss during the measurement. Higher focusing efficiency can be obtained if the objective lens can be replaced by the one with a higher NA. To further analyze the focal spot, its intensity profile along the horizontal direction is plotted in Fig. 3(g). The spot size (FWHM) is found to be 1.26 µm. This is close to the 3D-FDTD simulation result of 1.08 µm shown in Fig. 2(c). An ideal Airy function with the aperture diameter of 2 mm and the focal length of 1.76 mm is also plotted in Fig. 3(g). Compared with the measured profile, the Airy function has a narrower line shape. This might due to the fact that the intensity profile of the incident beam has a Gaussian intensity distribution, and hence the incident intensity at the edge of the metalens is weaker compare with the central region. It leads to the decrease of the effective lens diameter and NA, and consequently, the increase of the focal spot size. The image quality of a lens can be described by the modulation transfer function (MTF), which can be obtained by the

Fourier transform of the focal spot intensity [10,22]. The MTF of the metalens, together with the diffraction limit MTF, are presented in Fig. 3(h). The MTF curve of the metalens is close to the MTF curve of the diffraction limit at the lower frequency range. At the higher frequency, the metalens MTF curve drops faster than the diffraction limit MTF curve. It means the imaging quality of the object deteriorates faster with the metalens than that under the diffraction limit condition as the object frequency increases. This can be improved by optimizing the metalens design and the measurement system to achieve a focal spot close to the ideal Airy function.

In order to apply the metalens in an imaging system, the characterization setup in Fig. 3(e) is modified as shown in Fig. 4(a). An object is placed between the laser source and the metalens. The original objective lens is replaced by the one with 20× magnification and an NA of 0.4. The object is a sample with the text "940NM" on it. Below the text, there is a rectangular bar with the width of 100 µm. The image of the object is captured by the camera through the lens set in front of it. By controlling the relative position of the object and the metalens, the magnification of the image can be adjusted. The image of the text "940NM" with the 3× and 6.7× magnification are shown in Fig. 4(b) and 4(c), respectively. The clear text image at both magnifications further verifies the focusing functionality of the metalens.

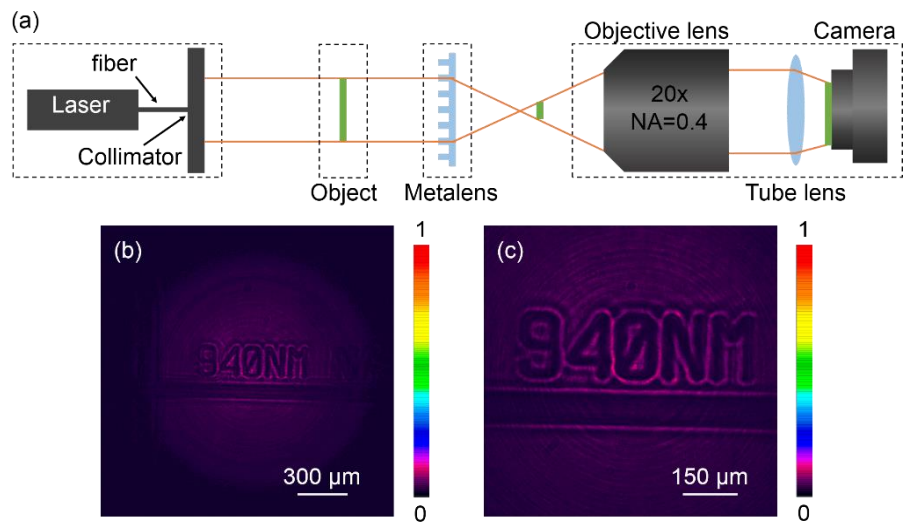

Fig. 4 (a) Schematic of the imaging system setup, including the metalens. Image taken for the text "940NM" with the (b) 3× and (c) 6.7× magnification.

## 4. Conclusion

The metalens working at 940 nm is designed by the 3D-FDTD simulation and fabricated on a 12-inch glass wafer using the CMOS-compatible technology. The NA and the focal spot size of the metalens are measured to be 0.494 and 1.26 µm, respectively, which matches well with the simulation. The focusing efficiency should be larger than the measured value of 29.2% since the NA of the objective lens used in the characterization setup is lower than that of the metalens. The metalens is applied in an imaging system, which further verifies its focusing functionality. The 193 nm ArF DUV immersion lithography and the rest fabrication process developed for this work can be used to fabricate the metalens working at even shorter wavelength, *i.e.* visible regime. This work provides a promising solution for the mass manufacturing of the metasurface based optical components on glass wafers.

## Acknowledgement

The authors acknowledge the RIE2020 Advanced Manufacturing and Engineering (AME) Domain's Core Funds: SERC Strategic Funds (A1818g0028).